\documentclass{article}
\usepackage{graphicx}
\usepackage{amsmath}

\newtheorem{theorem}{Theorem}

\newtheorem{definition}[theorem]{Definition}

\newtheorem{lemma}[theorem]{Lemma}

\date{}

\begin{document}

\title{Macrodimension - an invariant of local dynamics}
\author{V. A. Malyshev }
\maketitle

\begin{abstract}
We define a Markov process on the set of countable graphs with spins.
Transitions are local substitutions in the graph. It is proved that the
scaling macrodimension is an invariant of such dynamics.
\end{abstract}

\section{Introduction}

Process with a local interaction on a fixed lattice, graph or continuous
space, is now considered as one of the central objects in probability
theory. It was recognized recently that there exists a natural
generalization: where the graph itself is not fixed but randomly and locally
changes in time. Such processes appear in computer science (stochastic
grammars), in physics (quantum gravity) and in biology (DNA evolution).

General mathematical theory of such processes has started recently (see \cite
{m1, m2}). Weak convergence is an important tool for study of such processes.

The plan of the paper is the following. Markov dynamics is defined on the
set of finite graphs and countable graphs. For a countable graph the
macrodimension is defined. Roughly speaking the macrodimension of the graph
is $d$ if in the $N$-neighborhood of each vertex there is approximately $%
N^{d}$ vertices.

We prove that it is an invariant of any local dynamics (for any finite time)
under the following assumptions: dynamics should be local, locally bounded
and locally reversible. Note however that in the limit $t\rightarrow \infty $
the macrodimension can change.

\section{Local substitutions in spin graphs}

We consider non-directed connected graphs $G$ with the set of vertices $%
V=V(G)$ (finite or countable) and the set of links $L=L(G)$. The following
properties are always assumed: between each pair of vertices there is $1$ or 
$0$ edges; each node (vertex) has finite degree (the number of edges
incident to it). Denote $GF_{n}$ the set of all finite graphs with these
properties and where each node has degree not more than $n$. Denote $GC_{n}$
the set of all countable graphs with the same properties.

A \textit{subgraph} of $G$ is a subset $V_{1}\subset V$ of vertices together
with some links connecting pairs of vertices from $V_{1}$ and inherited from 
$L$. A \textit{regular subgraph} $G(V_{1})$ of $G$ is a subset $V_{1}\subset
V$ of vertices together with ALL links connecting pairs of vertices from $%
V_{1}$ and inherited from $L$.

The set $V$ of nodes is a metric space with the following metrics: the
distance $d(x,y)$ between vertices $x,y\in V$ is the minimum of the lengths
of paths connecting these vertices. The lengths of all edges are assumed to
be equal, say to some constant, assumed equal to $1$. Vertices connected by
a link are called neighbours. The neighbourhood (more exactly, $N$%
-neighbourhood) $O_{N}(v)$, of vertex $v$ in $G$ is the regular subgraph
with the set of vertices consisting of $v$ itself and of all vertices at
distance not greater than $N$ from $v$. Put $O_{1}(v)=O(v)$.

A \textit{spin graph } (also coloured graph, marked graph, spin system etc.)
is a pair $\alpha =(G,s)$, where $s=s(.)$ is a function $s:V\rightarrow S$
where $S$ is the set of ''spin values'', the alphabet. An isomorphism of
spin graphs is an isomorphism of graphs respecting the spins. The empty spin
graph $\emptyset $ is the empty graph with no spin. With some abuse of
notation we shall say also that $G$ is a spin graph and will call $\alpha $
simply a graph.

A spin graphs dynamics is a random sequence (where the moments $%
0<...<t_{k}<t_{k+1}<...,$ are also random) of spin graphs

\begin{equation*}
\alpha _{0}=(G_{0},s_{0}),\alpha _{t_{1}}=(G_{t_{1}},s_{t_{1}}),...,\alpha
_{t_{k}}=(G_{t_{k}},s_{t_{k}}),...
\end{equation*}
to be defined below. The graph $\alpha _{t_{k}}$ is obtained from $\alpha
_{t_{k-1}}$ by a simple transformation from some fixed class of
transformations.

\begin{definition}
The substitution rule (production) $Sub=(\Gamma ,\Gamma ^{\prime
},V_{0},\varphi )$ is defined by two ''small'' spin graphs $\Gamma $ and $%
\Gamma ^{\prime }$, subset $V_{0}\subset V=V(\Gamma )$ and mapping $\varphi
:V_{0}\rightarrow V^{\prime }=V(\Gamma ^{\prime })$; $V_{0}$ and $\Gamma
^{\prime }$ can be empty. The transformation (substituion) $T=T(Sub,\psi )$
of a spin graph $\alpha $, corresponding to a given substitution $Sub$ and
an isomorphism $\psi :\Gamma \rightarrow \Gamma _{1}$ onto a spin subgraph $%
\Gamma _{1}$ of $\alpha $, is defined in the following way. Consider
nonconnected union of $\alpha $ and $\Gamma ^{\prime }$, delete all links of 
$\Gamma _{1}$, delete all vertices of $\psi (V)\setminus \psi (V_{0})$
together with all links incident to them, identify each $\psi (v)\in \psi
(V_{0})$ with $v^{\prime }=\varphi (v)\in \Gamma ^{\prime }$. The function $s
$ on $V(G)\setminus V(\Gamma _{1})$ is inherited from $\alpha $ and on $%
V(\Gamma ^{\prime })$ - from $\Gamma ^{\prime }$. We denote the resulting
graph by $T(Sub,\psi )\alpha $.
\end{definition}

Examples of substitutions are: deleting a link, appending a link in a given
vertex with another new vertex, joining a pair of vertices (with no link
between them by a link, updating a spin value etc. Moreover the mere
possibility of these substituions can depend on the neighborhood of the
vertex where the substituion is to be done.

\begin{definition}
Graph grammar is defined by a finite set of substitutions $Sub_{i}=(\Gamma
_{i},\Gamma _{i}^{\prime },V_{i,0},\varphi _{i}),i=1,...,r$. We call a graph
grammar local if for all $i$ the $\Gamma _{i}$'s, corresponding to $Sub_{i}$%
, are connected. We call a graph grammar locally bounded if the subsets $%
GF_{n}$ and $GC_{n}$ are invariant with respect to all substitutions for
sufficiently large $n$.
\end{definition}

Let positive rates $\lambda _{i}=\lambda (Sub_{i}),i=1,...,r$, be given.
They define a continuous time Markov chain on $GF_{n}$, called random graph
grammar, in the following obvious way: at a given time interval $(t,t+dt)$
each possible transformation $T(Sub_{i},\psi )$ is produced with probability 
$\lambda _{i}dt+o(dt)$. It was proved in \cite{m1} that on $GF$ a minimal
Markov chain (non explosive) can be defined on all time interval $(0,\infty
) $.

\section{Macrodimension of a countable graph}

If one wants to consider an infinite graph as a topological space, this can
be done from local (micro) and global (macro) point of view. From micro
point of view graph is a 1-dimensional complex, thus having dimension 1. But
also a graph $G$ can be considered as a 1-dimensional skeleton of some
complex $C(G)$ of higher dimension. There are many ways to define $C(G)$ for
a given $G$. One of them is to consider the ''simplicial completion'' of the
1-dimensional complex $G$. That is the simplicial complex with vertices $V(G)
$, all edges are 1-dimensional simplices, a subset $S\subset V(G)$,
consisting of 3 vertices, is a simplex iff all its two element subsets are
simplices, etc. From macro point of view there are also many ways to define
its topological dimension, called in this case a macrodimension. We give
here one of the possible definitions. For any $x\in V(G)$ put 
\begin{equation*}
D_{n}(x)=\frac{\ln \left| O_{n}(x)\right| }{\ln n},\overline{D}(x)=\lim
\sup_{n\rightarrow \infty }D_{n}(x),\underline{D}(x)=\lim \inf_{n\rightarrow
\infty }D_{n}(x)
\end{equation*}

\begin{lemma}
(\cite{nore1}) For all $x,y$%
\begin{equation*}
\overline{D}(x)=\overline{D}(y)=\overline{D},\underline{D}(x)=\underline{D}%
(y)=\underline{D}
\end{equation*}
\end{lemma}

Proof. It follows easily from the fact that 
\begin{equation*}
O_{n-a}(y)\leq O_{n}(x)\leq O_{n+a}(y),a=\rho (x,y)
\end{equation*}

If $\overline{D}=\underline{D}=D_{S}$ then $D_{S}$ is called the scaling
macrodimension of $G$. For example, trees can have any positive scaling
dimension $1\leq D_{S}\leq \infty $. For the binary tree $D_{S}=\infty $ .
For any homogeneous lattice $L$ in the euclidean space $R^{d}$ the
macrodimension $D_{S}=d$. 

\section{Dynamics}

Let at time $t=0$ we have a connected spin graph $\alpha
(0)=(G=G(0),s(0))\in GC_{n}$ for some $n>0$ sufficiently large. We give now
exact definition of the dynamics $\alpha (t)=(G(t),s(t))$ on a small time
interval $(0,\varepsilon )$, see \cite{m2} for more details.

We need some notation. Fix some vertex in $G(0)$ and denote it $0$. Let $%
O_{N}=O_{N}(0)=G^{N}(0)$ be the $N$-neighbourhood of $0$ in $G$. Consider an
increasing family $O_{N}$ of finite spin graphs 
\begin{equation*}
O_{1}\subset ...\subset O_{N}\subset ...
\end{equation*}
Denote $\xi _{\beta }(t)$ the Markov process on $GF$ starting with finite
spin graph $\beta $ such that $\xi _{\beta }(0)=\beta $. We want to study
the limit $N\rightarrow \infty $ for the family $\xi ^{N}(t)=\xi _{O_{N}}(t)$
of Markov processes on the probability spaces $\Omega ^{N}$. Consider some
graph $G$ and a subset $V^{\prime }\subset V=V(G)$. Introduce the external
boundary

\begin{equation*}
\partial V^{\prime }=\left\{ v\in V\setminus V^{\prime }:\rho (v,V^{\prime
})=1\right\} 
\end{equation*}
A subset $V_{1}$ of vertices is connected if the regular subgraph with this
set of vertices is connected. We assume that the radii of all $\Gamma $
entering the definition of substitutions of our graph grammar do not exceed $%
1$. The latter assumption is only to do the presentation shorter.

Fix now $N$ and $t$. For the process $\xi ^{N}(t)$ and its trajectory $%
\omega $ we say that the vertex $v\in O_{N}$ has been touched on the time
interval $\left[ 0,t\right] $ if it belongs to one of the graphs $\psi
(\Gamma )$ in the random sequence of transformations on this time interval.
Denote $Q(N,t;\omega )$ the set of all vertices in $G^{N}(0)=O_{N}$ which
were touched by transformations of the process $\xi ^{N}$ during this time
interval. Introduce the random field $\eta ^{N}(v)$ on $V(G(0))$: $\eta
^{N}(v)=1$ if $v\in Q(N,t;\omega )$ and $\eta ^{N}(v)=0$ otherwise. For any
finite $T\subset V$ define the correlation functions 
\begin{equation*}
<\eta _{T}^{N}>=<\prod_{v\in T}\eta ^{N}(v)>
\end{equation*}
Limiting correlation functions (their existence was proved in \cite{m2})
define, by Kolmogorov theorem, a probability measure on $\left\{ 0,1\right\}
^{V(G(0))}$ (that is the limiting random field $\eta (v)$) or on the set of
all subsets of $V(G(0))$. Define for any initial graph $\alpha =\alpha (0)$
the random point set $E(\omega ,t,\alpha )=\left\{ v:\eta (v)=1\right\} $.

\begin{theorem}
There exists $t_{0}>0$ such that for any initial spin graph $\alpha $ and
any fixed $0\leq t<t_{0}$ the sequence of random fields $\eta ^{N}(v)$ tends
weakly to a random field $\eta (v)=\eta (v;\alpha )$ on $V(\alpha )$ when $%
N\rightarrow \infty $, that is $\eta _{T}^{N}>\rightarrow
<\eta _{T}>=<\prod_{v\in T}\eta (v)>$. Moreover, the random set $E(\omega
,t,\alpha )$ consists of countable number of finite connected components a.s.
\end{theorem}

Proof see in \cite{m2}.

We shall use the sets $E(\omega ,t,\alpha )$ for constructing dynamics on
the set of infinite graphs. For infinite dynamics the vertices of $%
V(G(0))\setminus E$ are interpreted as vertices which were touched by
substitutions on the time interval $[0,t]$. They stay unchanged together
with their spins.

Let us consider the probability space $\Omega _{1}=\left\{ 0,1\right\}
^{G(0)}$ where the limiting random field $\eta (v)$ is defined. Denote $\mu
_{1}$ the corresponding probability measure on $\Omega _{1}$. 

For given subgraph $B$ of $G(0)$ consider the process, starting with the
spin graph $B\cup \partial B$, that is the process $\xi _{B\cup \partial
B}(t)$ under the condition that all vertices in $B$ are touched but the
vertices on the boundary $\partial B$ of $B$ are not touched. Denote $\zeta
(t,B)$ this conditional process and let $\mu _{B}$ be the measure of this
process. Its trajectories on the time interval $\left[ 0,t\right] $ we
denote $\omega (B)$. Denote its probability space by $(\Omega _{B},\Sigma
_{B},\mu _{B})$.

For given $\omega _{1}$ let $B_{k}(\omega _{1}),k=1,2,...$, be all connected
components of $E(\omega ,t,\alpha )$. The limiting probability space for the
time interval $\left[ 0,t\right] $ is the set $\Omega $ of all arrays 
\begin{equation*}
\omega =\left( \omega _{1},\omega (B_{k}(\omega _{1})),k=1,2,...\right) 
\end{equation*}
Take some connected subsets $D_{1},...,D_{n}$ of $G(0)$ and some measurable
subsets $U_{i}\subset \Omega _{D_{i}}$. Let $C(U_{1},D_{1};...;U_{n},D_{n})$
the set of all $\omega $ such that each $D_{i}$ is equal to some of $%
B_{k}(\omega _{1})$ and moreover $\omega (B_{k}(\omega _{1}))\in U_{k}$.
Consider the minimal $\sigma $-algebra $\Sigma $ generated by all subsets $%
C(U_{1},D_{1};...;U_{n},D_{n})$ of $\Omega $.

The probability distribution $\mu $ is uniquely defined by the following
conditions: the projection of $\mu $ on $\left\{ 0,1\right\} ^{V}$ is equal
to $\mu _{1}$, and 
\begin{equation*}
\mu (C(U_{1},D_{1};...;U_{n},D_{n}))=\mu
_{1}(A(D_{1},...,D_{n}))\prod_{i=1}^{n}\mu _{D_{i}}(U_{i})
\end{equation*}
where $A(D_{1},...,D_{n})$ is the set of all $\omega _{1}$ such that $D_{i}$
are connected components of $E(\omega _{1},t,\alpha (0))$. Note that for
given $\omega _{1}$ the conditional distributions of trajectories on $%
B_{k}(\omega _{1})$ are independent for different $k$.

Points of the probability space $(\Omega ,\Sigma ,\mu )$ are, by definition,
the trajectories of the dynamics on the set of countable spingraphs. The
latter formula constitutes the cluster property. Similar representations can
be obtained for each finite $N$. In this sense the infinite dynamics is the
limit of the finite dynamics when $N\rightarrow \infty $. Note that the
interval $(0,\varepsilon )$ does not depend on the initial spin graph. Thus
by glueing together intervals $(n\varepsilon ,n\varepsilon +1)$ the dynamics
can be constructed on all time interval $(0,\infty )$. We can formulate now
the result.

\begin{theorem}
For any $t$ there exists dynamics on the set of countable spin graphs which
is the thermodynamic limit of the sequence of countable Markov chains.
Moreover it has the cluster property for $t<t_{0}$ uniformly in the initial
spin graphs.
\end{theorem}

\section{Dynamics and Metrics}

Here we define several classes of dynamics, accordingly to how they change
the metrical properties of a graph.

\subparagraph{Connectivity}

We assume that all substitutions of the fixed graph grammar respect
connectivity, that is if $\alpha $ is connected then $T(Sub_{i},\psi )\alpha 
$, for all possible $i$ and $\psi $, are also connected. It is easy to give
examples of local dynamics which does not respect connectivity. However, it
respects connectivity if together with locality we assume the following
state communicating condition on $GF$: if from some state $i$ one can reach
state $j$ in some steps with positive probability, then $i$ can be reached
from $j$ in some steps with positive probability (in particular one can
assume irreducibility).

\subparagraph{Metrical Boundedness}

\begin{lemma}
\label{L2}Assume that the graph grammar is local, locally bounded, respects
connectivity and that $G(0)\in GC_{n}$ for $n$ sufficiently large. Then it
is metrically bounded from above, that is there exists a constant $C=C(n)>0$
such that for each $G$ and each pair $x,y\in V(G)$ of vertices 
\begin{equation*}
d_{TG}(x,y)\leq d_{G}(x,y)+C
\end{equation*}
for each transformation $T=T(Sub_{i},\psi )$, where we write $TG$ instead of 
$T\alpha $.
\end{lemma}

Proof. Take some $C$ sufficiently large and assume that there exist $%
G,T=T(Sub_{i},\psi )$ and $x,y\in V(G)$ such that $d_{TG}(x,y)>d_{G}(x,y)+C$%
. We can assume that $x,y\in \partial \Gamma _{1}$. Take some $v\in \partial
\Gamma _{1}\subset V(G)$ and choose $N$ so that $0\ll n^{N}\ll C$. Then
applying $T$ to $O_{N}(v)$ we see that $T(O_{N}(v))$ will be nonconnected
because there is no paths from $x$ to $y$. In fact, the minimal path should
be longer than the number $n^{N}$ of vertices in $O_{N}(v)$.

\subparagraph{Local reversibility}

Important class of evolutions which respect connectivity comes from physics.
Consider a countable continuous time Markov chain with state space $X$ and
transition rates $\lambda _{ij},i,j\in X$. We can consider $X$ as the set of
vertices of the directed graph where there is a link from $i$ to $j$ iff $%
\lambda _{ij}$ is positive. It is called reversible (we do not assume
recurrence) if from $\lambda _{ij}>0$ it follows $\lambda _{ji}>0$ and, for
any cycle $\Gamma =(i_{1},i_{2}),...,(i_{n},i_{1})$, we have for $a_{ij}=%
\frac{\lambda _{ij}}{\lambda _{ji}}$ 
\begin{equation*}
a(\Gamma )=a_{i_{1}i_{2}}...a_{i_{n}i_{1}}=1,n\geq 2
\end{equation*}
Here $a_{ij}$ is a function on the set of links with values in $R\setminus
\left\{ 0\right\} $, $n$ is the length of the cycle. Reversible chaoin is
called locally reversible if, for some $n_{0}<\infty $ and for any finite $G$%
, the relations $a(\Gamma )=1$ for all $n$ follow from these relations for
all $n\leq n_{0}$. Let us give some examples of local reversibility:

\begin{enumerate}
\item  If all $\lambda _{ij}$ are positive then $n_{0}$ can be chosen equal
to $3$;

\item  Simple random walks in $Z^{d}$: $n_{0}=4$;

\item  Glauber dynamics for the Ising model in a finite volume: $n_{0}=4$.
\end{enumerate}

\section{Main Result}

\begin{theorem}
Assume the Markov chain on $GF$ to be local, locally bounded and locally
reversible. Then the scaling macrodimension is an invariant of the dynamics.
\end{theorem}

Proof. Let at time $t=0$ we have a connected spin graph $\alpha
(0)=(G=G(0),s(0))\in GC_{n}$ for some $n>0$ sufficiently large, having the
scaling macrodimension $d$. Consider the dynamics $\alpha (t)=(G(t),s(t))$
on a small time interval $(0,\varepsilon )$. It is sufficient to prove that
all $G(t)$ have the same scaling macrodimension $d$.

We need the following estimates.

\begin{lemma}
Consider the event that a cluster (connected component) $D$ of $E(\omega
,\varepsilon ,\alpha (0))$, containing a fixed vertex, has size $m$. Let $%
p(m)$ be the probability of this event. Then $p(m)<C\delta ^{m}$ for some $%
\delta <1$. Moreover, $\delta =\delta (\varepsilon )\rightarrow 0$ when $%
\varepsilon \rightarrow 0$.
\end{lemma}

This is the cluster estimate proved in \cite{m2}.

\begin{lemma}
Let $D$ be a cluster of size $m$ and $p(k,m,D),k>1$, be the probability that
the number of vertices in the graph $\zeta (\varepsilon ,D)$ becomes greater
than $km$. Then there exist constants $c>0$ and $\delta _{1}<1$ such that $%
p(k,m,D)<c\delta _{1}^{km}$ uniformly in $D$. Also $\delta _{1}=\delta
_{1}(\varepsilon )\rightarrow 0$ if $\varepsilon \rightarrow 0$.
\end{lemma}

Proof. Consider a kind of a pure growth process, more exactly the Markov
process on $Z_{+}$ with rates $\lambda i$ of jumping from $i$ to $i+r$. Let $%
q(r,m,k,t)$ be the probability that this process starting from $m$ particles
will have $km$ particles at time $t$. Then 
\begin{equation*}
q(1,1,k,t)=\exp (-\lambda t)(1-\exp (-\lambda t))^{k-1}
\end{equation*}
It follows 
\begin{equation*}
q(1,m,k,t)=\sum_{k_{1}+k_{2}+...+k_{m}=km}\prod_{i=1}^{m}\exp (-\lambda
t)(1-\exp (-\lambda t))^{k_{i}-1}<2^{km}(1-\exp (-\lambda t))^{m(k-1)}
\end{equation*}
Then for small $t$ 
\begin{equation*}
q(r,m,k,t)<2^{\frac{km}{r}}(1-\exp (-\lambda t))^{m(\frac{k}{r}-1)}<C\beta
^{km}
\end{equation*}
for $\beta =\beta (t)\rightarrow 0$ as $t\rightarrow 0$. At the same time,
see \cite{m1}, 
\begin{equation*}
p(k,m,D)\leq \sum_{j\geq k}q(r,m,j,t)
\end{equation*}
Consider the set $R(0)=V(G(0))\setminus E(\omega ,\varepsilon ,\alpha (0))$
of vertices which were not touched by the transformations on the time
interval $(0,\varepsilon )$, put $R_{N}(0)=R(0)\cap O_{N}(v_{0})$ where $%
v_{0}$ is a fixed vertex in $V(G(0))$. Let $v(\omega )\in R(0)$ be some
vertex on a minimal possible distance $r(\omega )$ from $0$. With some
ambiguity of notation we shall consider $v(\omega )$ and $R(0)$ belonging to 
$V(G(t))$ for all $t\in (0,\varepsilon )$. Let $O_{N}(t)=O_{N}(t,\omega )$
be the $N$-neighborhood of $v(\omega )$ in $V(G(t))$.

\begin{lemma}
There exists $C>1$ such that a.s. there exists $N_{0}=N_{0}(\omega )$ such
that for all $N>N_{0}$ we have 
\begin{equation*}
\left| O_{NC^{-1}}(\varepsilon ,\omega )\right| <\left| O_{N}(0)\right|
<\left| O_{NC}(\varepsilon ,\omega )\right| 
\end{equation*}
.\label{L1}.
\end{lemma}

Now the theorem easily follows. For example, 
\begin{equation*}
D_{S}(G)=\lim_{N\rightarrow \infty }\frac{\ln O_{N}(0)}{\ln N}=\lim \frac{%
\ln O_{N}(0)}{\ln (CN)}=\liminf \inf \frac{\ln O_{N}(0)}{\ln (CN)}\leq
\liminf \inf \frac{\ln O_{NC}(\epsilon )}{\ln (NC)}
\end{equation*}
and thus $D_{S}(G)\leq \underline{D}(G(\varepsilon ))$. Similarly, $%
D_{S}(G)\geq \overline{D}(G(\varepsilon ))$, and thus $D_{S}(G)=D_{S}(G(%
\varepsilon ))$.

Proof of the Lemma. Denote $d_{t}(x,y)$ the distance between points $x,y\in
V(G(t))$ in $G(t)$. Take any point $x\in R_{N}(0)$. Fix some $C$
sufficiently large and let $P(x,N)=P(x,N;C)$ be the probability of the event 
$A(x,N)$ that $x$ does not belong to $O_{NC}(t,\omega )$, that is $%
d_{t}(x,v(\omega ))>NC$. Then 
\begin{equation*}
P(x,N)<\beta _{1}^{NC}
\end{equation*}
In fact, compare $d_{0}(0,x)$ and $d_{t}(v(\omega ),x)$. Note first that 
\begin{equation*}
\Pr (\left| d_{0}(0,x)-d_{t}(v(\omega ),x)\right| >B)<const\beta ^{B}
\end{equation*}
for some $\beta $ small. Take some path $\Upsilon $ inside $O_{N}(0)$
between $x$ and $0$ of minimal length $l(\Upsilon )=d_{0}(0,x)\leq N$. Let $%
B(\Upsilon _{0}),\Upsilon _{0}\subset \Upsilon $, be the event that exactly
the vertices of $\Upsilon _{0}$ do not belong to $R(0)$. Let $D_{k}$ be all
clusters with which $\Upsilon $ intersects. Then the conditional probability 
\begin{equation*}
\Pr (\left| \cup _{k}D_{k}\right| >NC|\Upsilon _{0})<\beta ^{\sum_{k}\left|
D_{k}\setminus \Upsilon _{0}\right| }
\end{equation*}
for some $\beta <1$. Then 
\begin{equation*}
P(x,N)=\sum_{\Upsilon _{0}}\sum_{\left\{ D_{k}\right\} }\Pr (\Upsilon
_{0})\Pr (\left| \cup _{k}D_{k}\right| >NC|\Upsilon _{0})<\beta ^{\frac{NC}{2%
}}=\beta _{1}^{NC}
\end{equation*}
The probability $P(N)$ that $\left| O_{N}(0)\right| \geq \left|
O_{NC}(t,\omega )\right| $ is bounded by 
\begin{equation*}
P(N)\leq \sum_{x}P(x,N)\leq C(\gamma )N^{d+\gamma }\beta _{1}^{NC}
\end{equation*}
for any $\gamma >0$ and some $C(\gamma )>0$. Then the second inequality of
the Lemma follows from Borel-Cantelli Lemma. To prove the first inequality
we shall use the statement dual to Lemma \ref{L2}.

\begin{lemma}
Assume that the Markov chain on $GF$ to be local, locally bounded, locally
reversible and that $G(0)\in GC_{n}$ for $n$ sufficiently large. Then it is
metrically bounded from below, that is there exists a constant $C=C(n)>0$
such that for each $G$ and each pair $x,y\in V(G)$ of vertices 
\begin{equation*}
d_{TG}(x,y)\geq d_{G}(x,y)-C
\end{equation*}
for each transformation $T=T(Sub_{i},\psi )$, where we write $TG$ instead of 
$T\alpha $.
\end{lemma}

The first inequality of the Lemma \ref{L1}\ follows from this Lemma quite
similarly.

\end{document}